\documentclass[journal]{IEEEtran}
\usepackage[pdftex,final]{graphicx}
\usepackage{float} 
\usepackage{amsmath}                   
\usepackage{amsfonts}
\usepackage{amssymb}
\usepackage{color}

\ifCLASSINFOpdf
\else
\fi
\begin{document}
%
\title{Exploring Robustness of Cortical Morphometry in the presence of white matter lesions, using Diffusion Models for Lesion Filling}
%
%
%

\author{Vinzenz~Uhr, 
        Ivan Diaz,
        Christian Rummel, 
         and~Richard McKinley
}

%
%

\markboth{Journal of \LaTeX\ Class Files,~Vol.~14, No.~8, August~2015}%
{Shell \MakeLowercase{\textit{et al.}}: Leveraging Diffusion Models for Multiple Sclerosis Lesion Filling}
%



\maketitle

\begin{abstract}
Cortical thickness measurements from magnetic resonance imaging, an important biomarker in many neurodegenerative and neurological disorders, are derived by many tools from an initial voxel-wise tissue segmentation.  
White matter (WM) hypointensities in T1-weighted imaging, such as those arising from  multiple sclerosis or small vessel disease, are known to affect the output of brain segmentation methods and therefore bias cortical thickness measurements. These effects are well-documented among traditional brain segmentation tools but have not been studied extensively in tools based on deep-learning segmentations, which promise to be more robust.  In this paper, we explore the potential of deep learning to enhance the accuracy and efficiency of cortical thickness measurement in the presence of WM lesions, using a high-quality lesion filling algorithm leveraging denoising diffusion networks.  

A pseudo-3D U-Net architecture trained on the OASIS dataset to generate synthetic healthy tissue, conditioned on binary lesion masks derived from the MSSEG dataset, allows realistic removal of white matter lesions in multiple sclerosis patients.  By applying morphometry methods to patient images before and after lesion filling, we analysed robustness of  global and regional cortical thickness measurements in the presence of white matter lesions. Methods based on a  deep learning-based segmentation of the brain (Fastsurfer, DL+DiReCT, ANTsPyNet) exhibited greater robustness than those using classical segmentation methods (Freesurfer, ANTs).
\end{abstract}


%
\IEEEpeerreviewmaketitle

\section{Introduction}
%
%
%
%
\IEEEPARstart{W}{hite matter} (WM) lesions, often associated with neurological conditions like multiple sclerosis (MS), can significantly perturb tissue segmentation algorithms operating on magnetic resonance imaging (MRI), causing misclassification of tissue types. The misclassification varies considerably with lesion size and intensity, especially when the lesion intensity is similar to that of the Gray Matter (GM)/WM interface. As well as causing inaccuracies in volumetric gray mater measurements in brains with lesions, these perturbations can cause downstream biases in cortical thickness calculations. In the past different inpainting algorithms (also known as \emph{lesion filling}) have been proposed which can replace voxels within a lesion mask with white matter tissue intensities, leading to more robust measurements \cite{Battaglini2012EvaluatingMeasurements}, \cite{Valverde2014AMeasurements}, \cite{Amiri2018UrgentMRI}, \cite{Farazi2017InpaintingAtlas}, \cite{Almansour2021High-resolutionInpainting}.   
In recent years, deep learning has emerged as a powerful tool in medical image analysis, revolutionizing the field with its ability to automatically learn and extract meaningful features from large datasets. Deep learning techniques, especially convolutional neural networks (CNNs), have shown remarkable success in various medical imaging applications, including both brain segmentation \cite {XXX} and lesion filling \cite{Almansour2021High-resolutionInpainting}.  Deep-learning-based image segmentation models have been shown to dramatically outperform previous generations of brain segmentation tools.
Meanwhile denoising diffusion probabilistic models (DDPM) \cite{Ho2020DenoisingModels} have shown an impressive performance and experienced increasing popularity in medical image analysis \cite{Kazerouni2022DiffusionSurvey}.

In this article we use DDPMs to examine the robustness of DL and non-DL approaches to cortical morphometry.  Our contributions are
\begin{itemize}
    \item Evaluating different methods using noise diffusion models to inpaint WM lesions in MR images.
    \item Evaluating the impact of lesion filling on cortical thickness measurements using existing morphological tools, to determine their robustness to the presence of WM lesions.
\end{itemize}

\section{Background}
\subsection{Lesion Filling}
The presence of white matter lesions can significantly impact MR-based measurements like cortical thickness due to misclassification of different tissue types \cite{Battaglini2012EvaluatingMeasurements} \cite{Amiri2018UrgentMRI} \cite{Tiu2022PredictiveReview} \cite{Bieder2023Memory-EfficientProcessing} \cite{Magon2014WhiteStudy}. This misclassification is particularly problematic for WM lesions with size and intensity similar to the GM/WM interface and leads to overestimation of GM atrophy \cite{Gelineau-Morel2012TheSclerosis.}. Lesion filling algorithms have been developed to address this issue and improve measurements such as cortical volume, thickness and surface area estimation \cite{Amiri2018UrgentMRI}.

Early lesion filling approaches employed various strategies. For instance, \cite{Battaglini2012EvaluatingMeasurements} utilized lesion filling to enhance brain volume measurements, including normalized brain volume (NBV), normalized white matter volume (NWMV), normalized gray matter volume (NGMV), and percentage brain volume change (PBVC). Their method involved calculating intensity distributions of cerebrospinal fluid (CSF), CSF/GM, GM, and GM/WM from existing brain MR-images and filling WM lesions with pixel intensities randomly sampled from these distributions.

Another approach, proposed by \cite{Valverde2014AMeasurements}, involved refilling WM lesions by replacing lesion voxel intensities with random values drawn from a normal distribution based on the WM signal intensity of each two-dimensional slice. Segmentation of the slices was achieved using the fuzzy c-means algorithm.

Graph theoretical network analysis, a technique used to assess brain connectivity patterns, can also benefit from lesion filling. To reduce the variability in network analysis caused by WM lesions, \cite{vanderWeijden2022TheImaging} applied lesion filling by substituting lesion voxel intensities with intensities from nearby voxels. Their study suggests that lesion filling might improve the detection of network alterations in MS patients, but also highlights the potential for introducing artifacts. Therefore, caution is advised, especially for individuals with high lesion loads or lesions located at the WM/CSF or WM/GM interface.

More recent advancements leverage machine learning for lesion inpainting. \cite{Farazi2017InpaintingAtlas} employed a total variation model to improve registration performance with brain atlases. Inspired by Gated Convolution, \cite{Almansour2021High-resolutionInpainting} introduced a user-guided deep adversarial inpainting model capable of filling irregularly shaped holes in high-resolution T1w MR brain images. Training data generation involved synthesizing lesion masks by sampling and deforming random circles. Additional data augmentation techniques included rotation, cropping, flipping, noise addition, and varying brightness levels.

The emergence of DDPMs \cite{Ho2020DenoisingModels} offers a novel approach for high-quality image generation. DDPMs exhibit superior distribution coverage and training stability compared to adversarial loss-trained models, achieving state-of-the-art performance in various image synthesis tasks.

The International Brain Tumor Segmentation (BraTS) challenge in 2023 incorporated an inpainting challenge focused on synthesizing healthy brain tissue in glioma-affected regions \cite{Kofler2023TheInpainting}. Due to the high computational cost of 3D processing, \cite{Durrer2024DenoisingTissue} opted for a 2D diffusion model conditioned on glioma masks. While achieving comparable results to other participants, their approach resulted in stripe artifacts due to stacking of the 2D slices. Gaussian filtering was subsequently employed to mitigate these effects at the slice borders.

\cite{Almansour2021High-resolutionInpainting} addressed the high computational demands of 3D diffusion models by proposing several resource-reduction strategies. Notably, they introduced PatchDDM, a memory-efficient patch-based diffusion model that allows for inference on the entire volume while training solely on patches. Additional approaches included reducing self-attention layers, incorporating additive skip connections, and training on downsampled data.

In pursuit of improved inpainting quality for 3D MR-images, \cite{Durrer2024DenoisingInpainting} evaluated and modified various diffusion models, including 2D, pseudo-3D, and 3D models operating in image space, 3D wavelet or 3D latent space. Their findings suggest that the pseudo-3D model proposed by \cite{Zhu2023Make-A-Volume:Synthesis} achieved the best performance in terms of structural similarity index measure (SSIM), peak signal noise ratio (PSNR), and mean squared error (MSE). 
\subsection{Denoising Diffusion Probabilistic Models}
\label{sec:DDPM}
Diffusion models are a generative deep learning technique that leverage an approach for data synthesis. The core idea lies in progressively transforming a data sample $x_0$   from its original distribution into a sample $x_T$ from a normally distributed noise. The model then learns to progressively reverse this transformation process \cite{Sohl-Dickstein2015DeepThermodynamics}.

\begin{figure}[H]
    \centering
    $$q(x_t|x_{t-1})=\mathcal{N}(x_t;\sqrt{1-\beta_t}x_{t-1},\beta_tI)$$
    \includegraphics[width=0.9\linewidth]{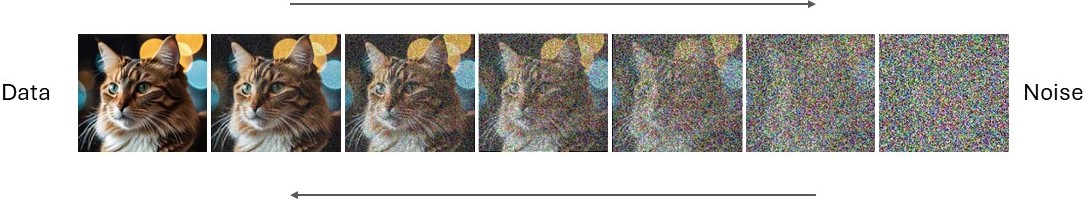}
    $$p_\theta(x_{t-1}|x_t)=\mathcal{N}(x_{t-1};\mu_\theta(x_t;t),\sigma^2_tI)$$
    \caption[Diffusion process]{Diffusion process from Data to Noise and reverse process from Noise to Data.}
    \label{fig:enter-label}
\end{figure}

The term forward process is the progressive transition from a clean image $x_0$ to pure noise $x_T$, via a series of steps, each adding additional random noise. At each step, zero-mean Gaussian noise is added, gradually increasing its strength until a maximum level is reached at a predefined endpoint $t=T$. This process is a \emph{Markov chain}, since the noisy image $x_t$ only depends on the immediately previous one $x_{t-1}$ (and not the whole sequence).  
The mathematical formulation behind this forward noising process $q$ is denoted by
\begin{equation}
    q(x_t|x_{t-1}):=\mathcal{N}(x_t;\sqrt{1-\beta_t}x_{t-1},\beta_tI) \label{equ:forwDiff_2}
\end{equation} 
Where $I$ represents the identity matrix and  $\beta_t$ the variance schedule, which controls the amount of noise added at each step based on the current step $t$.  The noisy image $x_t$ can be written 
\begin{multline} 
    x_t=\sqrt{\Bar{\alpha_t}}x_0+\sqrt{1-\Bar{\alpha_t}}\epsilon \\ \text{ where } \epsilon\sim\mathcal{N}(0,I) \text{ and } \alpha_t:=1-\beta_t \text{ and } \Bar{\alpha_t}=\prod_{s=1}^{t} \alpha_s \label{equ:forwDiff} 
\end{multline} 
The noise schedule $\beta_t$ is designed such that $\alpha_T\rightarrow0$ and $q(x_T|x_0)=\mathcal{N}(0,I)$.  Note that the Gaussian noise is used here for its mathematical properties (in particular, that the sum of two Gaussians is also a Gaussian), and does not imply any noise structure in the image $x_0$ (in particular, the use of Gaussian noise in the diffusion process is not incompatible with the fact that noise in MRI signals are not Gaussian)\\ \smallskip \\

The forward process provides training examples for the denoising process $p_\theta$, whose goal is to predict a less noisy sample $x_{t-1}$ from a noisy sample $x_t$. In general, the equation $q(x_{t-1}|x_t)\propto q(x_t|x_{t-1})q(x_{t-1})$ is intractable, but it can be approximated with a Gaussian for small transitions (small $\beta_t$). The equation for the reverse can be written as
\begin{equation}
    p_\theta(x_{t-1}|x_t)=\mathcal{N}(x_{t-1};\mu_\theta(x_t,t),\sigma_t^2I)
\end{equation}
The variance $\sigma_t^2$ can be fixed, eliminating the need to learn it explicitly. In practice it is easier to learn a model $\epsilon_\theta(x_t,t)$ which predicts the noise that needs to be removed at each step \cite{Ho2020DenoisingModels}. A sample $x_{t-1}$ can then be generated from $x_t$ by
\begin{equation} 
    x_{t-1}=\frac{1}{\sqrt{\alpha_t}}(x_t-\frac{1-\alpha_t}{\sqrt{1-\Bar\alpha_t}}\epsilon_\theta(x_t,t))+\sigma_tz \text{, with }z\sim\mathcal{N}(0,I)
\end{equation}
Given pure noise $x_T=\mathcal{N}(0,I)$, an image can be synthesized by iteratively applying equation 4 for all timesteps $t\in\{T,...,1\}$ to obtain the final prediction $x_0$. The diffusion model $\mu_\theta$ usually uses a U-Net architecture \cite{Ronneberger2015U-Net:Segmentation}. Training focuses on minimizing the MSE loss  
\begin{equation}
    \mathbb{E}_{x_0\sim q(x_0),\epsilon\sim\mathcal{N}(0,I)}[||\epsilon-\epsilon_\theta(x_t,t)||^2] \label{equ:loss}
\end{equation}  

\section{Methods}

\subsection{DDPMs for lesion filling}
CNNs operating on 3D volumes require significant GPU memory. To address this challenge, we treat the 3D volume as batches of 2D transversal slices and employ a 2D Unet for processing, stacking the results to yield a 3D image. \\
We explore two approaches in MR-images using diffusion models: conditional and unconditional. Both approaches utilize the ground truth MR-image $x$, a binary mask $m$ defining the lesion region, and the masked ground truth image $\hat{x}$.

\subsubsection{Conditional Model} \label{sec:CondFill}
\begin{figure}[H]
    \centering
    \includegraphics[width=0.9\linewidth]{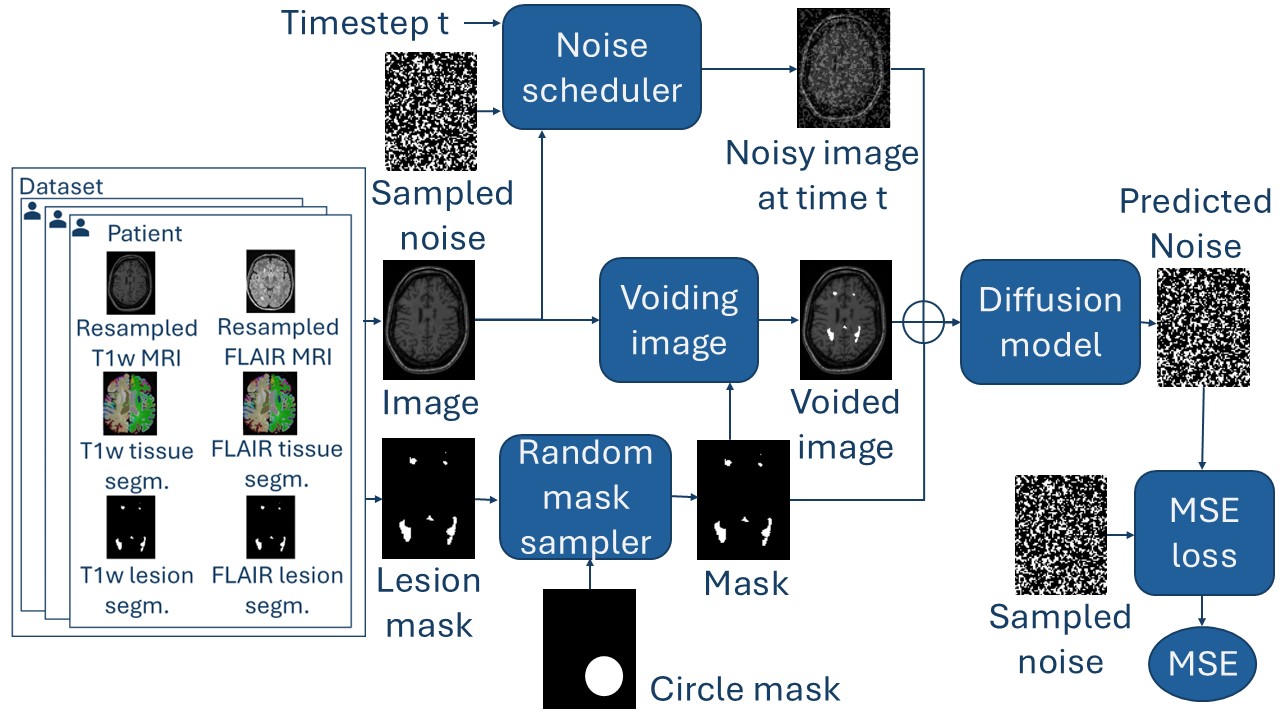}
    \caption[Training step involving the conditional mixture model]{Training step involving the conditional mixture model. An MR-image and its corresponding lesion mask are sampled from the dataset. Alternatively, with a 50\% probability, a random circle mask is sampled instead. This mask is used to void the portion of the image, which requires inpainting. Additionally, a random timestep $t$ and random noise matching the image shape are sampled. These are used to generate a noisy image as described in Section \ref{sec:DDPM}. The mask, the noisy image, and the voided image are concatenated and fed into the diffusion model, which aims to predict the sampled noise. The predicted and sampled noise are used to calculate the MSE.}
    \label{fig:trainingstep}
\end{figure}

The conditional approach trains a diffusion model conditioned on the masked ground truth image and the binary mask. The conditioning information is incorporated through channel-wise concatenation. At each timestep $t$ during reverse diffusion, the model receives the concatenated input of the noisy image $x_t$, the masked ground truth image $\hat{x}$ and the binary mask $m$. The objective is to predict the noise term for calculating a less noisy image. This leads to the loss function, 
\begin{equation}
    \mathbb{E}_{x_0\sim q(x_0),\epsilon\sim\mathcal{N}(0,I)}[||\epsilon-\epsilon_\theta(((\sqrt{\Bar{\alpha_t}}x_0+\sqrt{1-\Bar{\alpha_t}}\epsilon)\oplus\hat{x}\oplus m),t)||^2]
\end{equation} 
For sampling, we employ Denoising Diffusion Implicit Models (DDIM) \cite{Song2021DENOISINGMODELS}, a computationally efficient class of iterative probabilistic models that share the training procedure of DDPM. DDIM utilizes a non-Markovian sampling process, which is deterministic.
The sequence of a training step for the conditional model is described later in Figure \ref{fig:trainingstep}.

\subsubsection{Unconditional Model}
The unconditional approach does not use conditioning information during training. We train an unconditional DDPM as a generative prior, as described in Section \ref{sec:DDPM}. This essentially creates a model that can produce random 2D brain MRI samples. To condition the generation process, we modify the reverse diffusion iterations by sampling masked regions using the provided image information, as proposed in the RePaint paper \cite{Lugmayr2022RePaint:Models}. This technique does not modify the original DDPM network and is applicable to any inpainting mask distribution.

Inpainting aims to predict missing pixels within a masked region based on surrounding image information. Each reverse step from image $x_t$ (noisy) to $x_{t-1}$ (less noisy) depends solely on the noisy image $x_t$, which consists of unknown pixels within the mask $m\odot x_t$  and known pixels outside the mask $(1-m)\odot x_t$. The known pixels can be calculated for each timestep based on the forward process (Equation \ref{equ:forwDiff}). The RePaint approach proposes separate sampling processes for unknown and known pixels during the reverse step, resulting in the following expression:
\begin{equation}
    x^{known}_{t-1}\sim\mathcal{N}(\sqrt{\Bar{\alpha_t}}x_0,(1-\Bar{\alpha_t})I)
\end{equation}
\begin{equation}
    x^{unknown}_{t-1}\sim\mathcal{N}(\mu_\theta(x_t,t),\Sigma_\theta(x_t,t))
\end{equation}
\begin{equation}
    x_{t-1}=(1-m)\odot x^{known}_{t-1}+m\odot x^{unknown}_{t-1}
\end{equation}
Here, $x_{t-1}^{known}$ is sampled using the known pixels in the given image $(1-m)\odot x_t$, while $x_{t-1}^{unknown}$ is sampled from the model given the previous iteration $x_t$. These are then combined to form the new sample $x_{t-1}$  using the mask $m$.

\subsubsection{U-Net Architectures}
As model we used a 2D U-Net and a pseudo-3D U-Net \cite{Zhu2023Make-A-Volume:Synthesis} which achieved high scores in another study comparing different diffusion models architectures for 3D healthy brain inpainting \cite{Durrer2024DenoisingInpainting}. 
The 2D U-Net has an architecture similar to \cite{Nichol2021ImprovedModels}. It uses six feature map resolutions with two convolutional residual blocks per resolution level and one self-attention block at the 16x16 resolution after each convolutional block. From highest to lowest resolution the U-Net stages use (128, 128, 256, 256, 512, 512) channels. 

The pseudo-3D U-Net has, in addition to the 2D U-Net, a volumetric layer inside the residual block after each 2D convolution. Pseudo-3D convolutions result from 1D convolutions in the z-axis, requiring the batch to be rearranged before and after. Following \cite{Durrer2024DenoisingInpainting} we apply the model in the image space and directly use the pseudo-3D convolutions without the proposed fine-tuning strategy used by the original paper \cite{Zhu2023Make-A-Volume:Synthesis}.
To setup the U-Net models and the training environment, we used the python library diffusers from huggingface \cite{vonPlatenDiffusers:Models}.

\subsection{Datasets}
To train the models, a dataset was created by combining healthy subjects from the OASIS project \cite{Marcus2007OpenAdults} with MS patients from the MICCAI challenge \cite{Commowick2021MultipleDataset}.

The OASIS dataset consisted of T1w MRI scans from 20 healthy subjects. It was divided into training and validation sets, with 16 samples used for training and 4 for validation.

The MICCAI challenge dataset comprised of MRI scans of 15 patients diagnosed with MS. Each scan included both T1w and FLAIR images, with the FLAIR images containing manually segmented lesion masks. This dataset was also split into training and validation sets, with 13 samples for training and 2 for validation. 

Section \ref{sec:MaskGeneration} explores the use of additional synthetic masks in the form of random circle masks. To evaluate their effectiveness on larger datasets, the BraTS Inpainting Challenge 2023 dataset \cite{Kofler2023TheInpainting} was employed. The training set of this dataset comprises 1251 brains. Since the challenge has concluded and online analysis of the validation set is no longer possible, the training set was divided into 90\% training data and 10\% test data.

The impact of lesion filling on cortical thickness measurements was conducted using a test set composed of 65 patients diagnosed with RR-MS.  This data originated from an internal longitudinal study conducted at the Insel hospital. All patients had been undergoing Natalizumab treatment for over two years and had at least four MRI scans performed over a period of approximately six months each, with corresponding clinical evaluations.  MRI scans included a combination of 1.5T and 3T datasets with a slice thickness of 1mm or less in the T1w sequences.  For each patient, the T1w and FLAIR images from their final visit, typically containing the highest lesion burden, were used for testing.

\subsubsection{Preprocessing}
All T1w images undergo resampling to a standardized size of 256x256x256 voxels with a 1.0x1.0x1.0 mm voxel size and are reoriented to RAS orientation. FLAIR images are resampled to 160x256x256 voxels. The resampling process is carried out using nibabel.processing.conform \cite{Brett2023Nipy/nibabel:5.2.0}. Values below 0.01 are discarded as noise and the remaining data is scaled to the range [-1, 1]. A deep learning-based tissue segmentation is performed on the T1w images for each patient using the DL+DiReCT model \cite{Rebsamen2020DirectParcellation}. To accelerate this process, the parallelization program GNU Parallel \cite{Tange2018GNU2018} is employed. For datasets with existing lesion masks, these are registered from FLAIR to T1w images using NiftyReg \cite{Modat2010FastUnits}. In the absence of lesion masks in the test set, a separate segmentation model DeepSCAN \cite{McKinley2021SimultaneousNetworks} is utilized to identify MS lesions. Only 2D slices containing WM, based on the DL+DiReCT segmentation, were incorporated for training. 
\begin{figure}[H]
    \centering
    \includegraphics[width=0.9\linewidth]{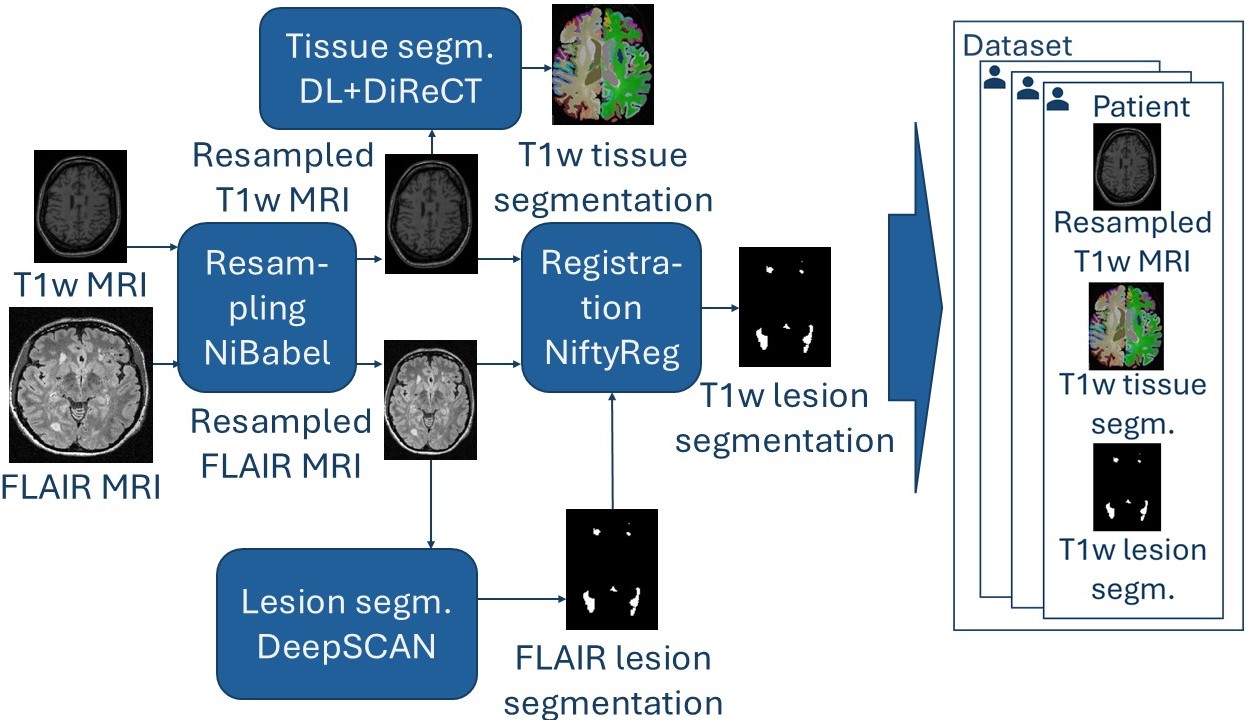}
    \caption{Creation of the training dataset}
    \label{fig:enter-label}
\end{figure}

\subsubsection{Mask Generation} \label{sec:MaskGeneration}
The distribution of masks employed for training a conditional model can have an impact on the performance of the model \cite{Suvorov2022Resolution-robustConvolutions} \cite{Yu2019Free-formConvolution} \cite{Zeng2020High-ResolutionUpsampling} \cite{Liu2018ImageConvolutions}. Conditional models were trained on the healthy subject images using lesion masks obtained from the MS patients. To achieve this, each lesion mask from the MS patients was registered to every T1w image from the healthy subjects. This resulted in 15 registered lesion masks for each of the 20 healthy patients. 

Each lesion mask was restricted to WM tissue by multiplying it with a binary WM mask derived from the DL+DiReCT segmentation.
To augment the diversity of the masks, the set of connected lesions was computed for each mask. During training, a different set of connected lesions was sampled and used as the lesion mask.

Given the limited dataset of masks, a secondary approach was explored that utilized a second mask distribution consisting of random circle masks with varying locations and sizes. This resulted in three models being trained: One model trained on the distribution of real lesion masks (conditional lesions model), one model trained on the distribution of random circle masks (conditional circles model) and one model trained with a combination of 50\% real lesion masks and 50\% random circle masks (conditional mixture model). 

\subsubsection{Min-SNR Loss weighting} \label{sec:minSNR}
During the training of the unconditional model, the validation loss indicated faster overfitting with smaller timesteps (e.g., 200 steps) compared to larger ones (e.g., 1000 steps). This could be due to the different levels of difficulty inherent in the time steps of the diffusion models. Predicting added noise becomes progressively easier as the image approaches pure noise. Consequently, bigger timesteps naturally result in lower MSE and correspondingly weaker gradients compared to smaller timesteps. This imbalance leads the training process to prioritize optimization for smaller timesteps.

To address this and achieve a more balanced loss function, we explored the min-SNR weighting strategy proposed in \cite{Hang2023EfficientStrategy}. This approach advocates for adapting loss weights assigned to individual timesteps based on clamped signal noise ratios.

\subsection{Evaluation of the DDPM}
\subsubsection{Metrics}
During the training, the model was evaluated at regular intervals using the validation dataset. MSE, PSNR, and SSIM \cite{Wang2004ImageSimilarity} are calculated inside the masks, outside the masks, and across the entire image. LPIPS \cite{Zhang2018TheMetric} is solely evaluated over the whole image

All metrics are measured on 2D images. The model version with the highest SSIM score is periodically saved to disc. 

\subsubsection{Mask Dilation}
The evaluation revealed the presence of artifacts at the boundaries of inpainted lesions. This occurs because RePaint replaces all areas outside the designated mask with the original image.  If the annotated masks don't fully encompass the entire lesion, these small residual areas of the original lesion can lead to border artifacts.  Conditional models exhibited similar, though less pronounced, artifacts.
To address this issue, we implemented a minor, one-pixel dilation restricted to WM regions. This dilation strongly minimizes the artifacts.

\subsection{Evaluation of the robustness of cortical thickness methods}
Having identified the best performing lesion filling method, further evaluations centers on the influence of lesion filling on cortical thickness measurements. Cortical thickness assessments are performed on sixty-five patients from the test set, both before and after lesion filling, using five methods: ANTs \cite{Tustison2014Large-scaleMeasurements}, ANTsPyNet \cite{Tustison2021TheImaging}, FreeSurfer \cite{Fischl2012FreeSurfer}, FastSurfer \cite{Henschel2020FastSurferPipeline} and DL+DiReCT \cite{Rebsamen2020DirectParcellation}. FreeSurfer calculates cortical thickness by modeling the cortical band as a surface mesh. FastSurfer, a more recent method, replicates FreeSurfer's anatomical segmentation, including surface reconstruction, but leverages deep learning techniques to accelerate the process. On the other hand, ANTs,ANTSPyNet and DL+DiReCT all build on cortical thickness method based on diffeomorphic registration-based cortical thickness (DiReCT) applied to an atlas-based segmentation. ANTsPyNet extends this approach by incorporating deep learning for segmentation: ANTs applys this to a Bayesian EM-based segmentation, while ANTsPyNet and DL+DiReCT both use a deep-learning algorithm to provide the segmentation.   To assess robustness, we apply each method before and after lesion filling.   \cite{Jovicich2013BrainSegmentations}.
The average absolute changes relative to the mean (\%) are calculated using the following formula:
$$\epsilon_\mu=\frac{100}{N}\sum^N_{i=1}\frac{1}{2}\sum^2_{t=1}\frac{|m_{i,t}-\mu_i|}{\mu_i}$$
Where N is the number of patients, $m_1$ the measurement before lesion filling, $m_2$ the measurement after lesion filling and $\mu_i=\frac{1}{2}\sum^2_{t=1}m_{i,t}$ the within-patient mean. 
This calculation is performed both for the global mean thickness (averaging the mean thickness of the left and right hemispheres) and across anatomical regions defined by the Desikan-Killiany (DK) atlas \cite{Desikan2006AnInterest}. Since ANTs and ANTsPyNet do not deliver regional statistics, for those methods we average over the parcellation derived from DL+DiReCT.  In a subanalysis, we excluded cases with MS lesions located near the cortical surface (juxtacortical lesions) which might lead to lesion-filling errors: to identify patients with juxtacortical lesions, the binary lesion masks are dilated by one pixel and multiplied with the tissue segmentations. Patients with lesions outside WM are excluded in the second analysis.

\section{Results}
\subsection{Performance of the lesion-filling models}
The 3D conditional model trained with a balanced mixture of lesion masks and random circle masks emerges as the top-performing model, attaining a SSIM of 0.96 and LPIPS of 2e-4 on the evaluation set. Metrics measured during training can be viewed in Appendix \ref{chap:AppendixA}. A comparative analysis between 2D and pseudo-3D models reveals that the latter consistently outperforms the former across all metrics. Furthermore, within the realm of conditional models, the architecture trained with random circle masks demonstrates superior performance compared to its lesion mask-trained counterpart.
\begin{table}[hbt!]
    \centering
    \begin{tabular}{|c|c|c|c|c|}
         \hline
         & SSIM & PSNR & MSE & LPIPS \\ 
         \hline
         2D unconditional RePaint & 0.83 & 28 & 8.2e-3 & 2.0e-3 \\
         \hline
         2D conditional circles & 0.9 & 32 & 4e-3 & 2e-3  \\
         \hline
         2D conditional lesions & 0.85 & 28 & 0.01 & 5e-3 \\
         \hline
         2D conditional mixture & 0.9 & 33 & 4e-3 & 1e-3  \\
         \hline
         3D unconditional RePaint & 0.90 & 32 & 3e-3 & 9e-4  \\
         \hline
         3D conditional circles & 0.95 & 38 & 1e-3 & 3e-4  \\
         \hline
         3D conditional lesions & 0.93 & 34 & 3e-3 & 4e-4  \\
         \hline
         \textbf{3D conditional mixture} & \textbf{0.96} & \textbf{39} & \textbf{8e-4} & \textbf{2e-4}  \\
         \hline
    \end{tabular}
    \caption[Metrics measured with validation dataset]{Metrics measured with validation dataset. SSIM, PSNR and MSE are measured inside the mask and LPIPS over the full image.}
    \label{tab:metricsFilling}
\end{table}
\begin{figure}[H]
    \centering
    \includegraphics[width=0.8\linewidth]{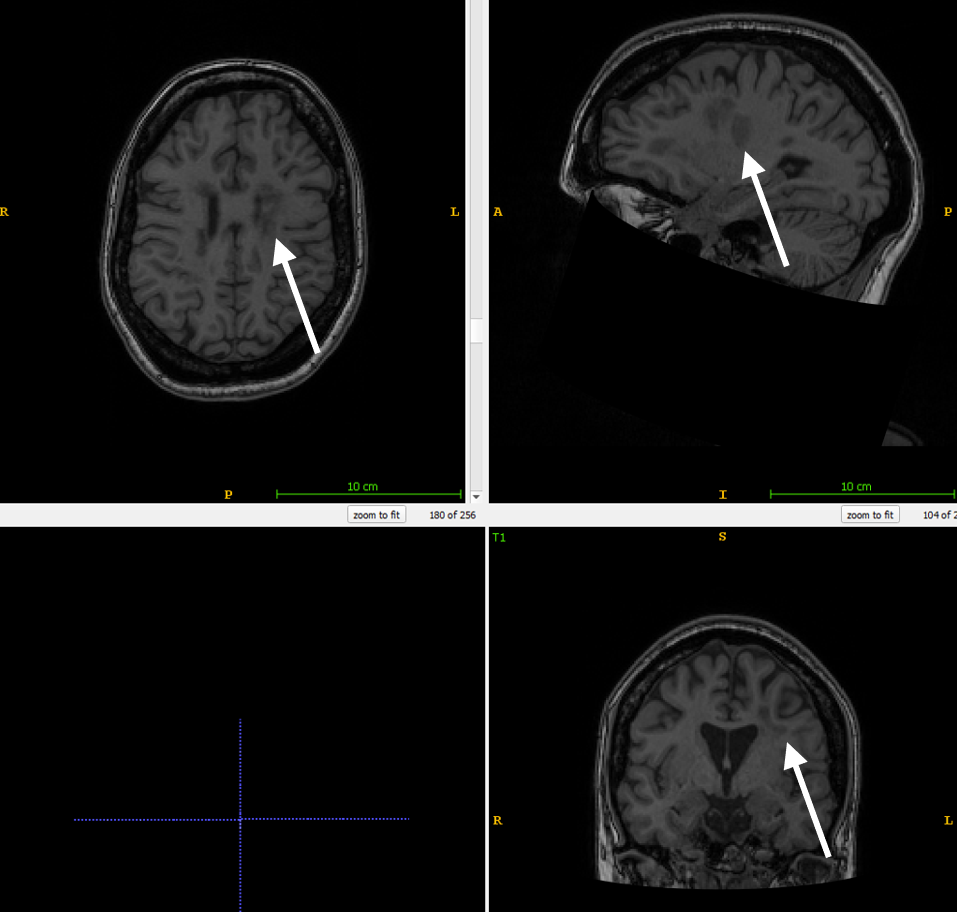}
    \caption{T1w before lesion filling with conditional mixture model}
    \label{fig:enter-label}
\end{figure}
\begin{figure}[H]
    \centering
    \includegraphics[width=0.8\linewidth]{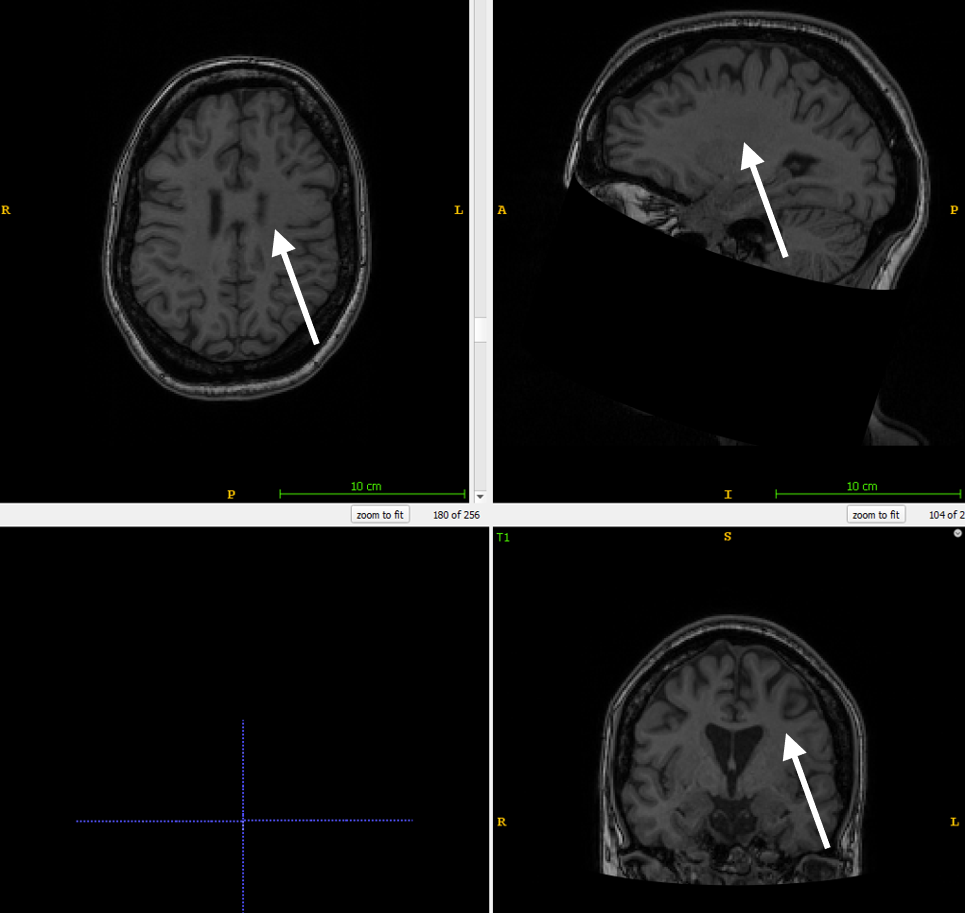}
    \caption{T1w after lesion filling with conditional mixture model}
    \label{fig:enter-label}
\end{figure}

A significant difference exists in terms of inference time. Due to the resampling approach, the inference time for the unconditional RePaint model is substantially longer compared to the conditional models. For a batch of 16 samples on a single Nvidia RTX A6000 40GB GPU, the inference time is 45 seconds for the conditional mixture model and 350 seconds for the unconditional RePaint model.


\subsection{Robustness of Cortical Thickness methods}
Table \ref{tab:reprErrors} presents mean reproducibility errors for both global mean thickness and the average across all 68 ROIs, calculated using data from 65 patients. To account for potential influences of juxtacortical lesions, we excluded patients with such lesions and recalculated the same measurements for the remaining 17 patients, with results displayed in Table \ref{tab:reprErrorsNoJuxta}. Lesion filling was performed using the 3D conditional mixture model.
ANTsPyNet and FastSurfer, which incorporate deep learning, show significantly improved robustness compared to their predecessors. Furthermore, the DL+DiReCT approach yields a substantial additional reduction in error.

\begin{figure}[H]
    \centering
    \includegraphics[width=0.8\linewidth]{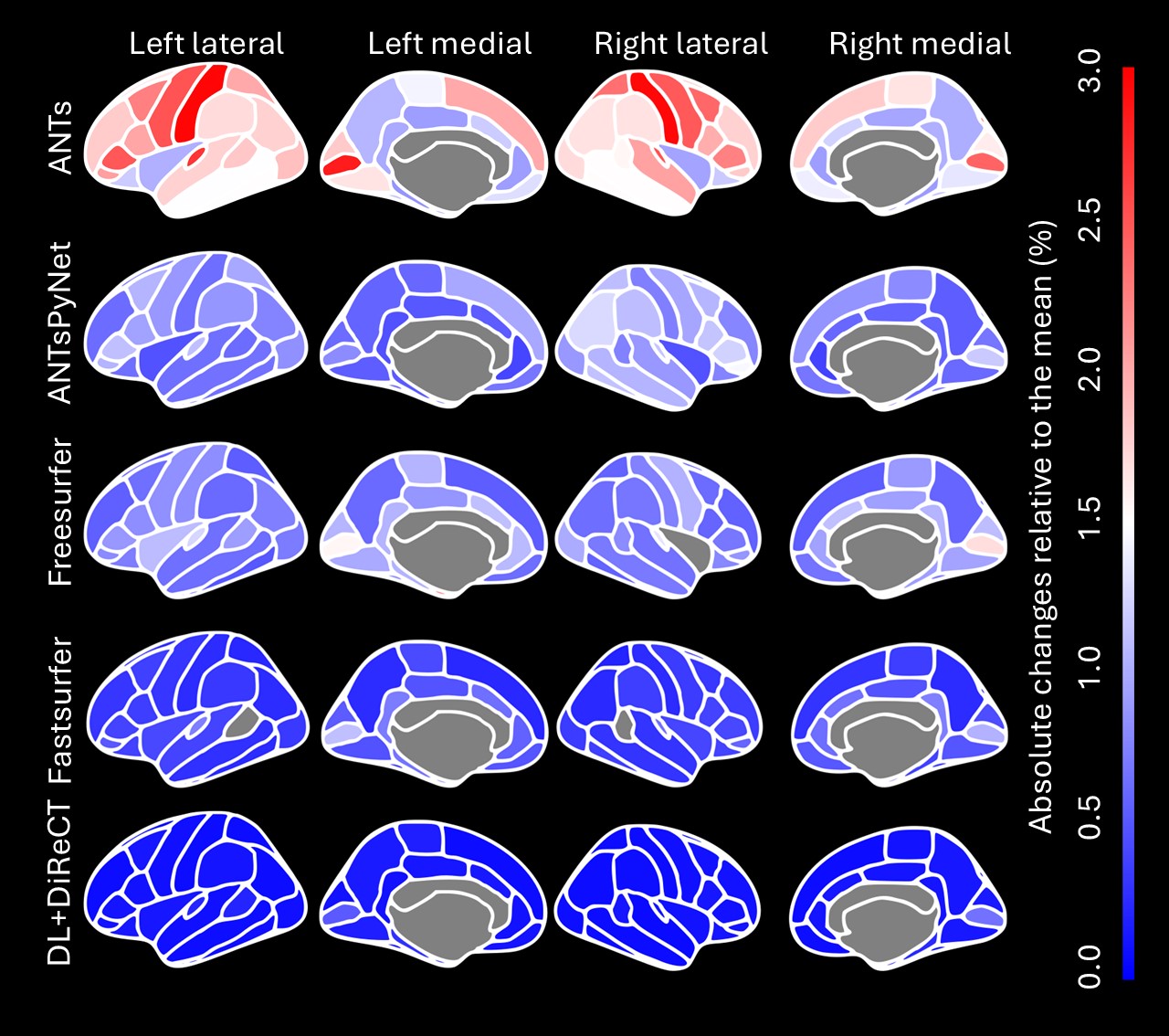}
    \caption[Color-coded reproducibility errors]{Color-coded reproducibility errors of the ROI-wise average cortical thicknesses evaluated on all samples.}
    \label{fig:reproducibilityErrorROI}
\end{figure}

Comparing robustness across different regions reveals consistent superiority of the newer deep learning-based methods (see Figure \ref{fig:reproducibilityErrorROI}). 
ANTsPyNet's least robust region is the left frontal pole with a 1.4\% error. In contrast, FastSurfer's least robust regions are the left and right pericalcarine regions, with errors of 1.1\% and 1.0\% respectively. 
Similarly, DL+DiReCT exhibits the lowest robustness in the right and left pericalcarine regions, with error rates of 0.6\% and 0.5\%, respectively.
\begin{table}[hbt!]
    \centering
    \begin{tabular}{|c|c|c|}
     \hline 
     & Global mean thickness (\%) & ROI-average (\%) \\
     \hline
     ANTs & 1.31 & 1.68 \\
     \hline
     ANTsPyNet & 0.52 & 0.84 \\
     \hline
     FreeSurfer & 0.51 & 0.92 \\
     \hline
     FastSurfer & 0.14 & 0.45 \\
     \hline
     \textbf{DL+DiReCT} & \textbf{0.05} & \textbf{0.14} \\
     \hline 
    \end{tabular}
    \caption{Mean reproducibility errors}
    \label{tab:reprErrors}
\end{table}
\begin{table}[hbt!]
    \centering
    \begin{tabular}{|c|c|c|} 
     \hline 
     & Global mean thickness (\%) & ROI-average (\%) \\
     \hline
     ANTs & 1.34 & 1.68 \\
     \hline
     ANTsPyNet & 0.38 & 0.81 \\
     \hline
     FreeSurfer & 0.61 & 0.90 \\
     \hline
     FastSurfer & 0.13 & 0.43 \\
     \hline
     \textbf{DL+DiReCT} & \textbf{0.04} & \textbf{0.12} \\
     \hline 
    \end{tabular}
    \caption{Mean reproducibility errors without patients with juxtacortical lesions}
    \label{tab:reprErrorsNoJuxta}
\end{table}

\section{Discussion}

Across the whole cortex and regionally, the impact of lesion filling on cortical thickness measurements varies significantly depending on the morphometric tool employed. While the original ANTs model is strongly affected, the newer ANTsPyNet model, incorporating deep learning, shows a much smaller impact. Similarly, the newer FastSurfer model, also incorporating deep learning, improves the error rate compared to its predecessor, FreeSurfer. DL+DiReCT exhibits the smallest differences among all models. These findings suggest that deep learning models are more robust to WM lesions than classical methods.  

When considering the lesion filling models, we can observe some general trends: conditional models perform better than unconditional models, and 3D models perform better than 2D models.  The better performance is intuitive, as conditional models are explicitly trained for inpainting, while unconditional models are reused through the RePaint sample approach.   The conditional models have significantly longer training times (days versus hours), but this is more than offset by faster image generation at test time.


Comparing different training regimes, we found that training with additional circle masks enhanced performance. Surprisingly, training exclusively with random circle masks yielded better results than using only lesion masks, suggesting that a broader range of unrealistic masks distributed across MR-images is beneficial within the given dataset, rather than a smaller set of masks sampled from the true mask distribution. 


\section{Conclusion}
We successfully developed a deep learning model for filling MS lesions in MR-images, using it to observe the superior robustness of morphometric pipelines based on deep learning segmentations. This raises the possibility that lesion filling might become obsolete with the increasing adoption of more modern tools: meanwhile, researchers preferring to remain with more established tools may find it useful to perform lesion filling before analysis.

\subsection{Limitations}
The dataset used in this study is relatively small, which might limit the representativeness of the ethnic groups included. 
The current study focused on filling multiple lesions simultaneously. The performance of the model for inpainting single lesions while preserving others remains unexplored and could potentially differ.
We utilized lesion masks created by doctors based on their interpretation of MR-images. It's important to note that lesions can also influence the surrounding brain tissue, which may not be readily identifiable by humans on current MR-images. The extent of this influence and its relevance for lesion filling is a separate research question and may vary depending on the specific use case.

\subsection{Outlook}
The models developed for filling MS lesions, could be applied to other inpainting tasks. However, performance might vary, especially considering the training objectives. It would be interesting to determine if the performance advantage of conditional models over unconditional ones persists in these new applications. Improving the unconditional model to match the performance of the conditional model is another potential area of research. This is desirable due to the unconditional model’s shorter training time and independence from mask distribution.

Although we demonstrated that deep learning-based tools are more robust to MS lesions than older methods when measuring cortical thickness, a larger population study is necessary to definitively establish the obsolescence of lesion filling.

\ifCLASSOPTIONcaptionsoff
  \newpage
\fi



\bibliographystyle{IEEEtran}
\bibliography{references}
%

%

\appendices
\section{Quantitative and Qualitative Training Progression}
\label{chap:AppendixA}
This chapter presents both quantitative and qualitative visualizations of the different model's training progression. 

\begin{figure}[H]
    \centering
    \includegraphics[width=0.9\linewidth]{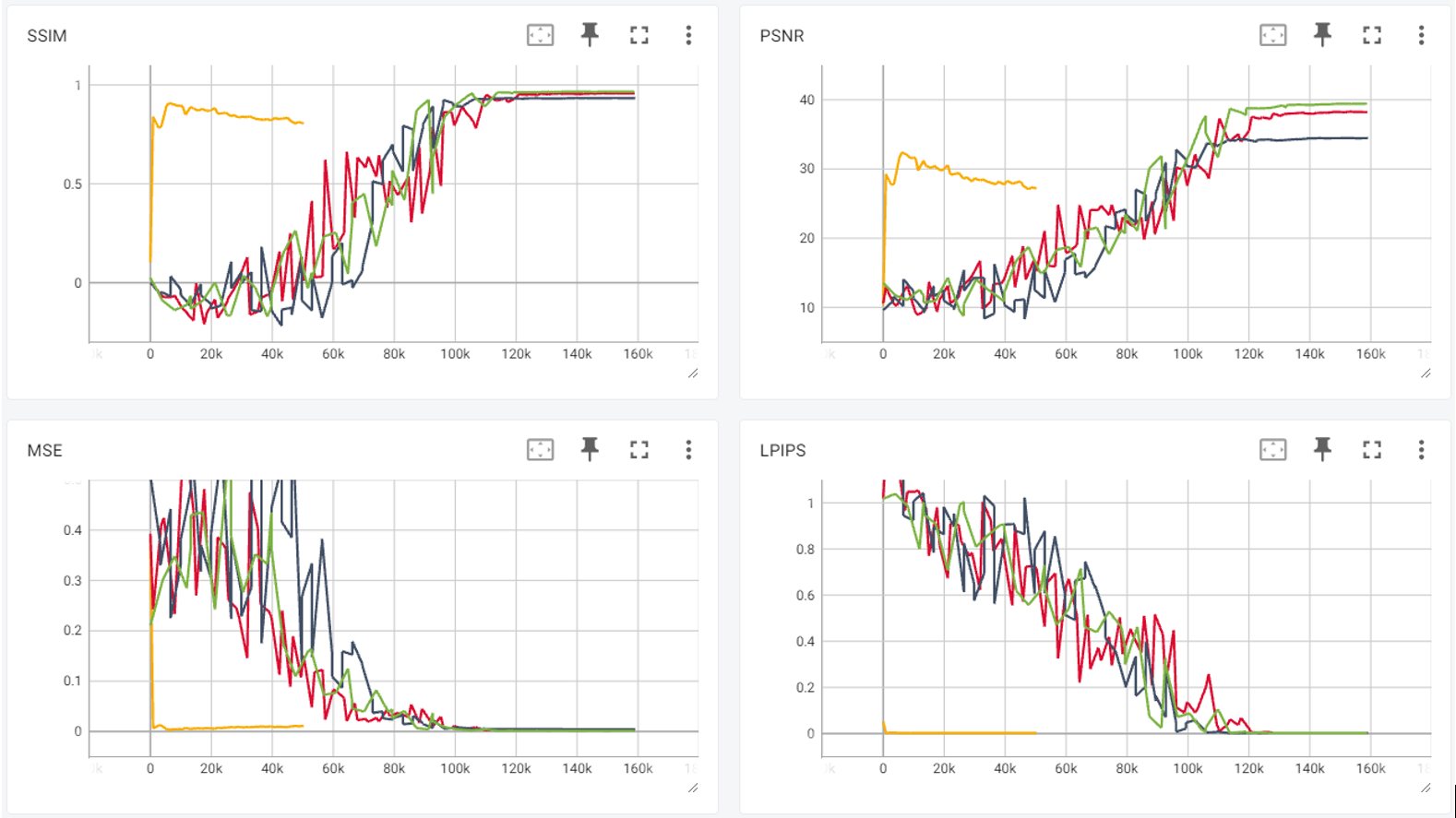}
    \caption[Training metrics of different 3D model’s]{Training metrics of the 3D model’s conditional mixture (green), conditional circles (red), conditional lesions (black) and unconditional RePaint (orange).}
    \label{fig:enter-label}
\end{figure}
\begin{figure}[H]
    \centering
    \includegraphics[width=0.9\linewidth]{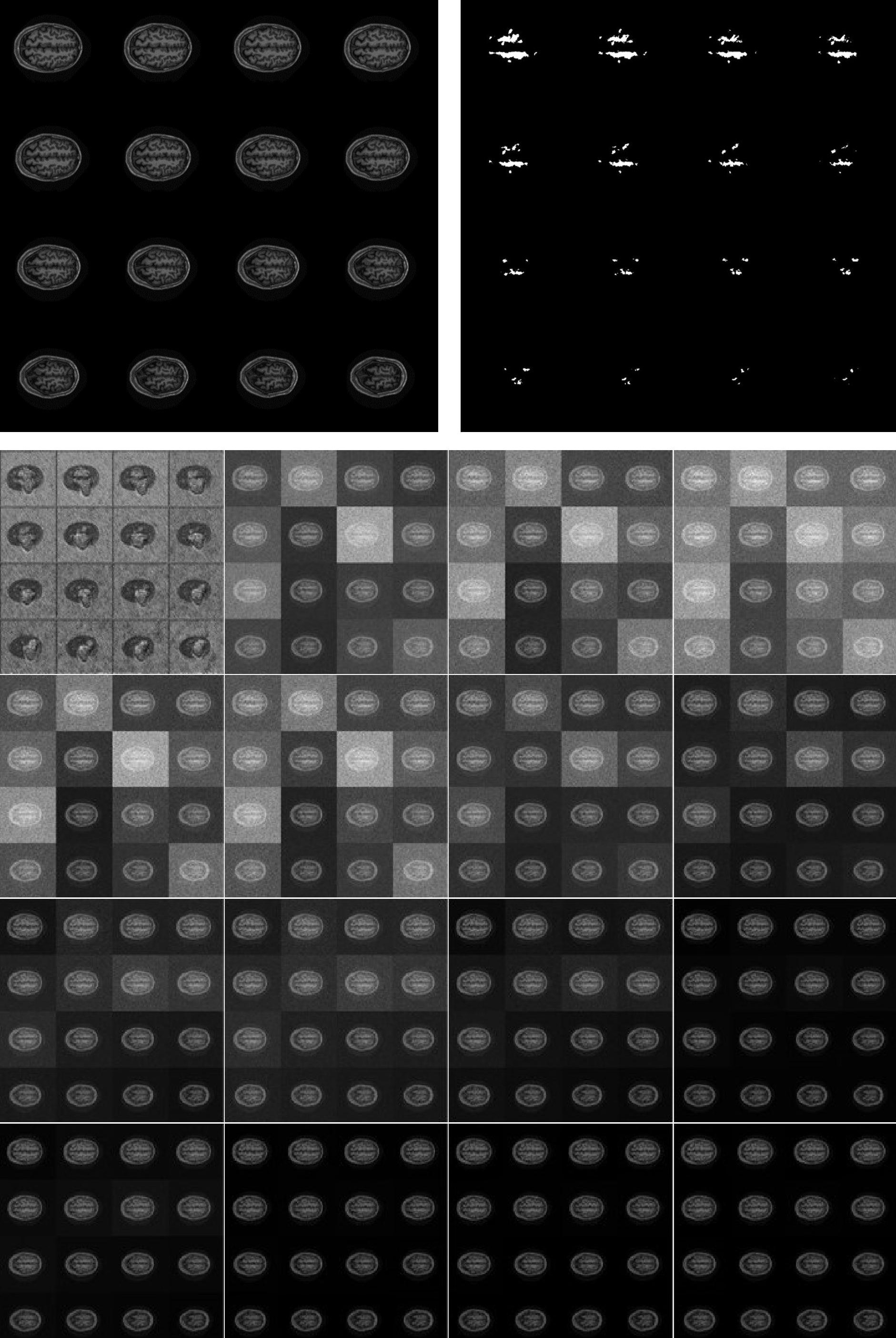}
    \caption[16 image-mask pairs for evaluation]{16 image-mask pairs for evaluation at the top right and left corners (Top). Below, the results of the 3D conditional mixture model training in a grid ordered from left to right and top to bottom at specific timesteps: 000001, 011’914, 017’210, 026’476, 034’419, 039’715, 051’628, 056’924, 066’190, 074’133, 079’429, 091’342, 096’638, 105’904, 113’847, 119’143}
    \label{fig:enter-label}
\end{figure}
\begin{figure}
    \centering
    \includegraphics[width=0.9\linewidth]{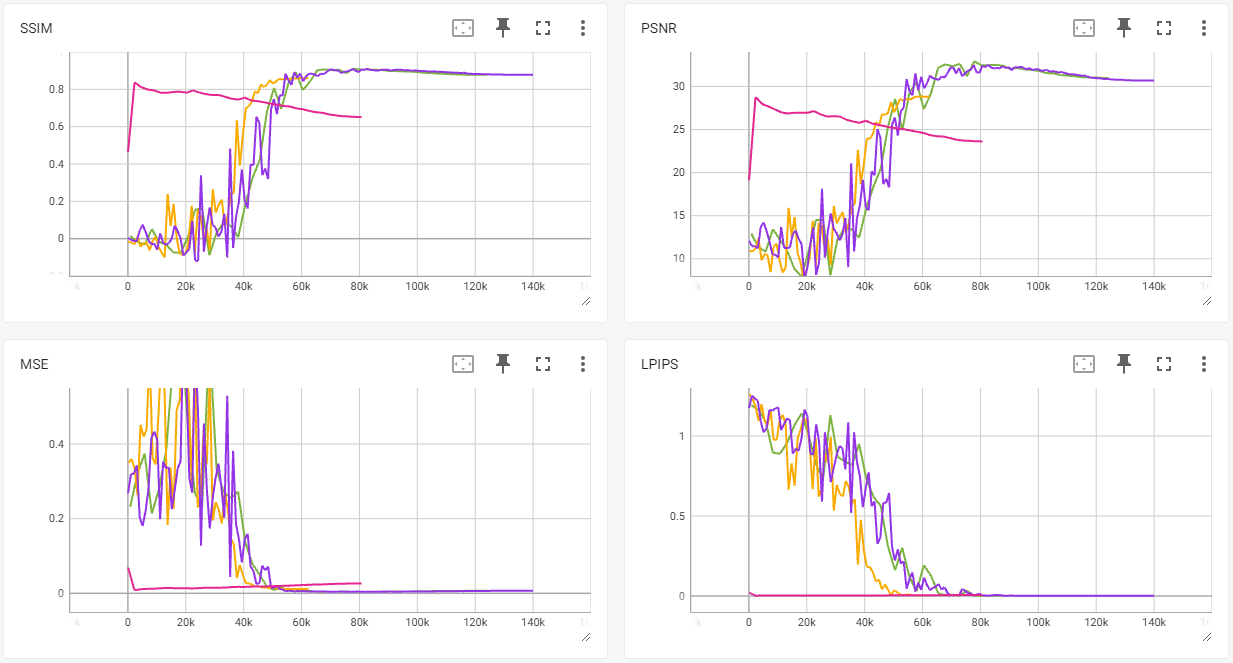}
    \caption{Training metrics of the 2D model’s conditional mixture (green), conditional circles (purple), conditional lesions (orange) and unconditional RePaint (red). }
    \label{fig:enter-label}
\end{figure} 

\begin{figure}[H]
    \centering
    \includegraphics[width=1\linewidth]{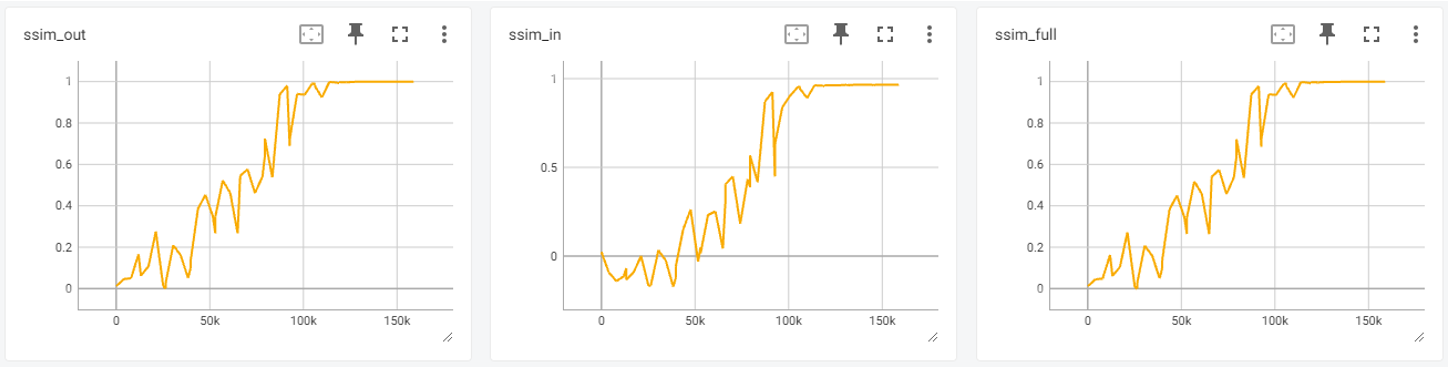}
    \caption[SSIM metric outside and inside the mask]{SSIM metric of the 3D conditional mixture model outside and inside the mask and over the entire image.}
    \label{fig:enter-label}
\end{figure} 

\section{Training Environment}
\begin{table}[hbt!]
    \centering 
    \begin{tabular}{ |c|c| } 
         \hline
         Number of training diffusion steps & 1000 \\ 
         \hline
         Number of inference steps & 50 \\ 
         \hline
         Batch size & 16 \\ 
         \hline
         Learning rate & 1e-4 \\ 
         \hline
         Optimizer & AdamW \\ 
         \hline
         Learning rate scheduler & Cosine with 500 steps warmup \\ 
         \hline
         RePaint Jump length & 8 \\ 
         \hline
         RePaint Resample & 10 \\ 
         \hline
    \end{tabular}
    \caption{Hyperparameters for training and evaluation}
    \label{tab:Hyperparam} 
\end{table} 

 \begin{table}[hbt!]
    \centering 
    \begin{tabular}{ |c|c| } 
         \hline 
         GPU	& 3x Nvidia RTX A6000 40GB \\ 
         \hline 
         CPU	& 64x Intel Xeon Gold 6226R @ 2.9Ghz \\ 
         \hline 
         RAM	& 196 GB \\ 
         \hline  
    \end{tabular}
    \caption{Hardware}
    \label{tab:Hardware} 
\end{table}
\subsection{Artifacts XXXappendix?}
Incomplete lesion masking leads to recognizable artifacts in the form of residual borders at the lesion edges. This phenomenon is particularly pronounced in the unconditional RePaint approach, which replaces regions outside the mask with original image content. Although less obvious, conditional models also exhibit similar artifacts. A small one-pixel dilation limited to the WM effectively mitigates this problem.

The intrinsic two-dimensional nature of 2D models leads to another artifact: inconsistencies along the z-axis, manifesting as visible stripes. The pseudo-3D models successfully mitigate these z-axis irregularities. 
\begin{figure}[H]
    \centering
    \includegraphics[width=0.8\linewidth]{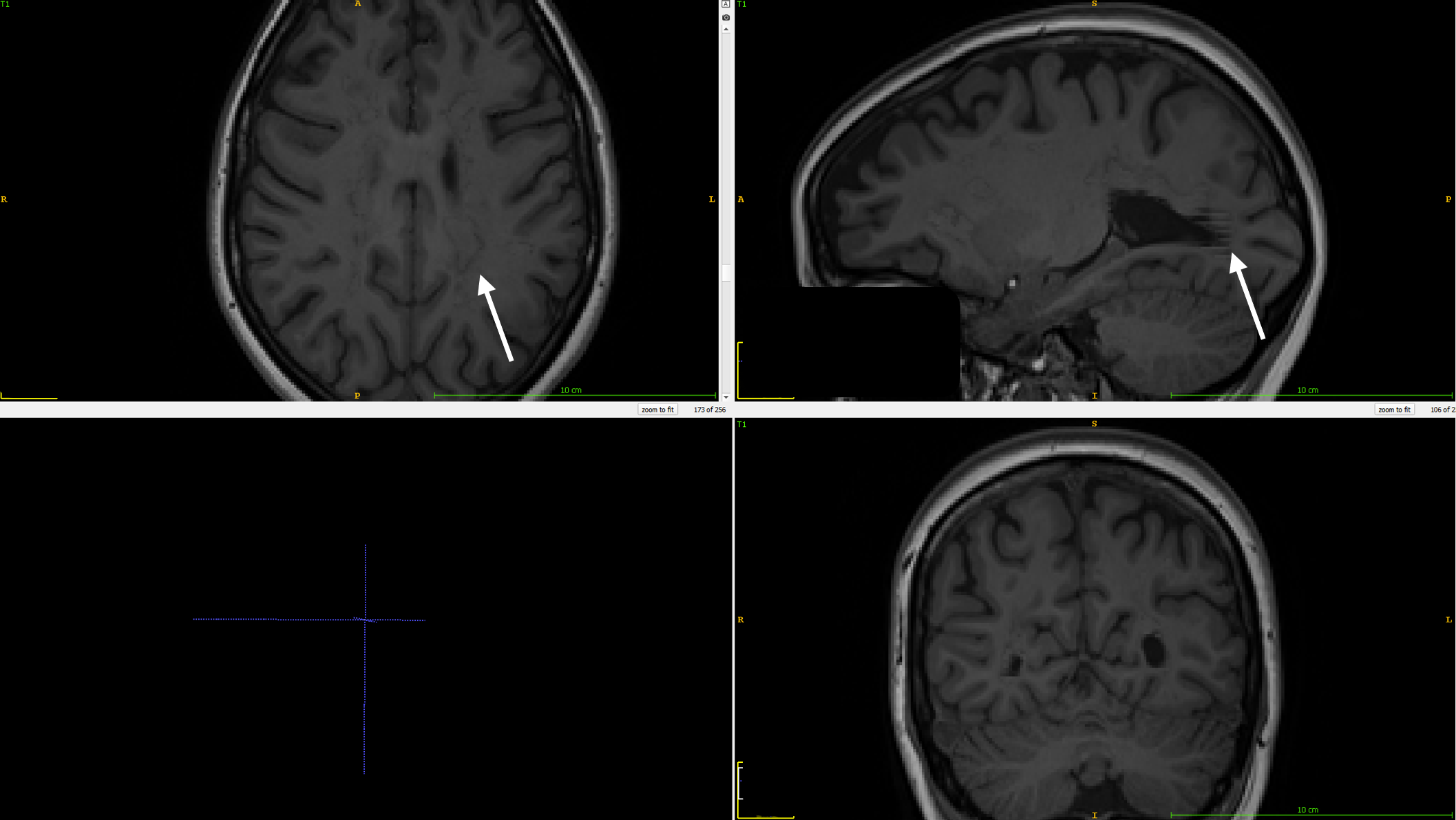}
    \caption[Border and stripe artifacts]{Border (left arrow) and stripe artifacts (right arrow)}
    \label{fig:enter-label}
\end{figure}

\subsection{Training Duration XXXapendix?}
The unconditional RePaint model exhibits significantly faster convergence compared to conditional models, achieving a peak SSIM of 0.9 after only 6000 training steps, while conditional models require approximately 90,000 steps to reach comparable performance.
\begin{figure}[H]
    \centering
    \includegraphics[width=0.9\linewidth]{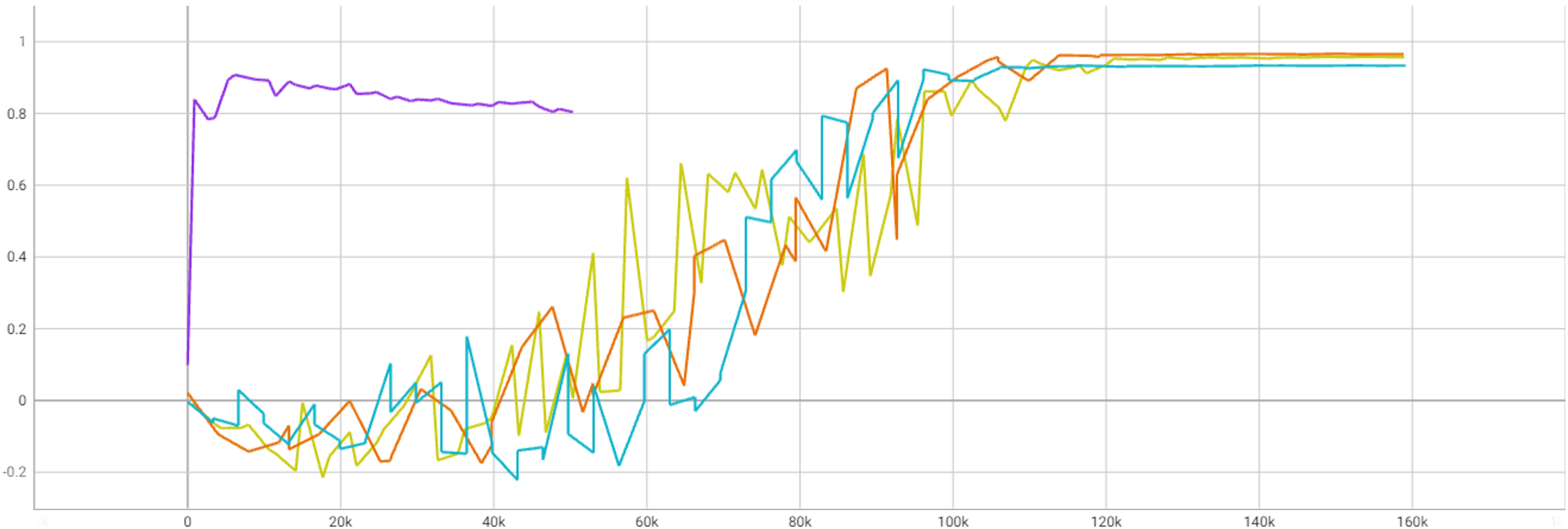}
    \caption[SSIM score during training]{SSIM score during training of the 4 3D models unconditional RePaint (violet), conditional mixture (red), conditional circles (yellow) and conditional lesions (blue).}
    \label{fig:enter-label}
\end{figure}
Interestingly, RePaint achieves optimal performance when the underlying unconditional model remains unconverged. This phenomenon is evident when sampling random 2D images using a DDIM sampler instead of the RePaint sampler, resulting in highly noisy outputs. While RePaint's strong guidance produces high-quality inpainting results, DDIM sampling reveals the underlying unconditional model's immaturity. This raises the question of whether preventing overfitting and refining the unconditional RePaint model can match or surpass the performance of conditional models while requiring substantially less training time. 
\begin{figure}[H]
    \centering
    \includegraphics[width=0.5\linewidth]{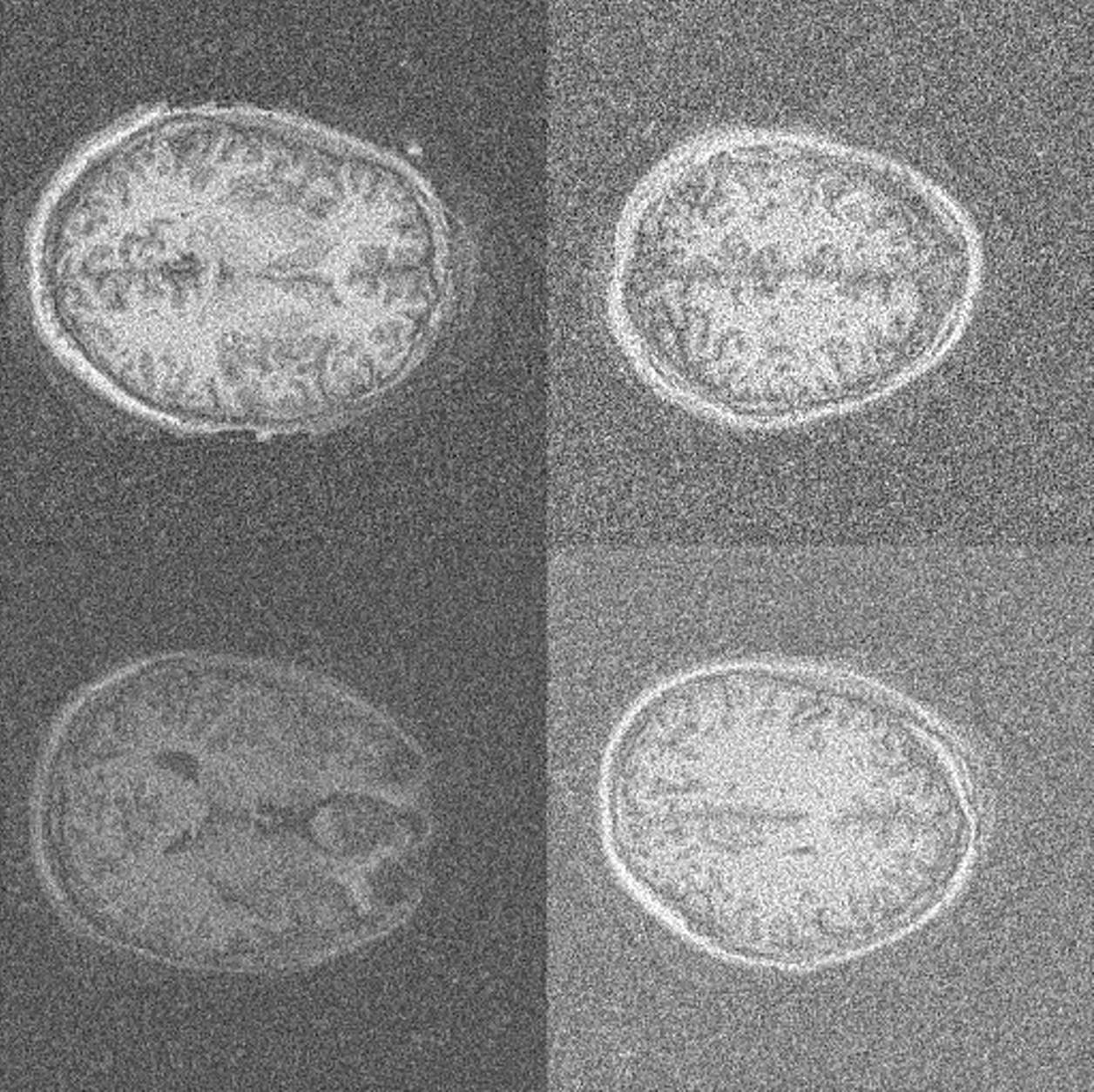}
    \caption[Samples of unconditional model with DDIM sampler]{Samples of unconditional model with DDIM sampler at a timestep in training, where the model achieves its best scores with the RePaint sampler.}
    \label{fig:enter-label}
\end{figure}
Examining the validation loss per timestep reveals that smaller timesteps begin to overfit while larger timesteps continue learning. To counteract this imbalance, we adopted the min-SNR weighting strategy outlined in Section \ref{sec:minSNR}.
\begin{figure}[H]
    \centering
    \includegraphics[width=0.9\linewidth]{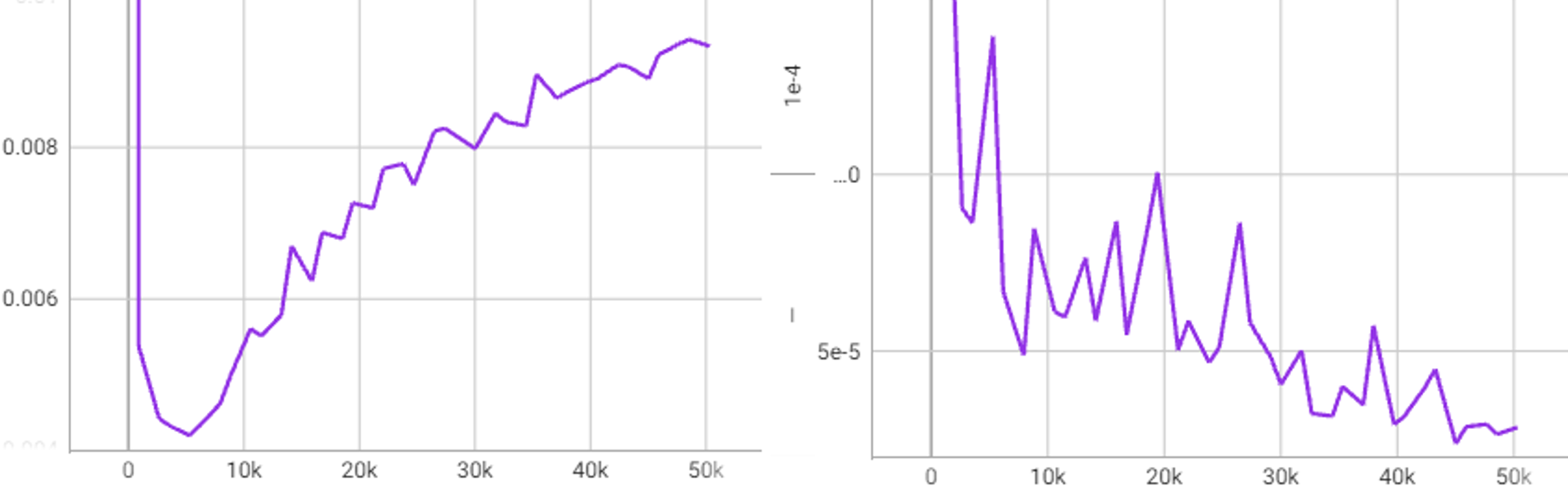}
    \caption[Comparison of the validation loss]{Comparison of the validation loss of timestep 200 (left) and 800 (right).}
    \label{fig:comparisonValLoss}
\end{figure}
\begin{table}[hbt!]
    \centering
    \begin{tabular}{|c|c|c|c|c|}
         \hline
         & SSIM & PSNR & MSE & LPIPS \\
         \hline 
         2D Unconditional RePaint & 0.83 & 28 & 8.2e-3 & \textbf{2.0e-3} \\
         \hline
         2D Uncon. RePaint min-SNR & 0.83 & \textbf{29} & \textbf{7.3e-3} & 2.8e-3 \\
         \hline
    \end{tabular}
    \caption[Min-SNR metrics measured with validation dataset]{Min-SNR metrics measured with validation dataset. SSIM, PSNR and MSE are measured inside the mask and LPIPS over the full image.}
    \label{tab:metricMinSNR}
\end{table}
Min-SNR loss mitigated overfitting, reducing the overall validation loss across all timesteps to 0.008 compared to 0.013 with unweighted MSE loss. However, metrics such as SSIM, PSNR, MSE, and LPIPS showed no significant improvement. To simplify the training process, min-SNR loss was excluded from the model training.
%








\end{document}